\documentclass[prb,twocolumn,preprintnumbers,amsmath,amssymb]{revtex4}
\usepackage{graphicx}
\usepackage{epsfig}
\usepackage{amsmath,amscd}
\usepackage{psfrag}
\usepackage[dvipdfm]{hyperref}
\usepackage{amsmath,amscd}
\usepackage{psfrag}

\begin{document}


\title{Maximum screening fields of superconducting multilayer structures.} 

\author{Alex Gurevich} 
\affiliation{Department of Physics and Center for Accelerator Science, Old Dominion University, Norfolk, Virginia 23529, USA}
\email{gurevich@odu.edu}

\date{\today}

\begin{abstract}

It is shown that a multilayer comprised of alternating thin superconducting and insulating layers on a thick substrate can fully screen the applied magnetic field exceeding the superheating fields $H_s$ of both the superconducting layers and the substrate, the maximum Meissner field is achieved at an optimum multilayer thickness. For instance, a dirty layer of thickness $\sim 0.1\; \mu$m at the Nb surface could increase $H_s\simeq 240$ mT of a clean Nb up to $H_s\simeq 290$ mT. Optimized multilayers of Nb$_3$Sn, NbN, some of the iron pnictides, or alloyed Nb deposited onto the surface of the Nb resonator cavities could potentially double the rf breakdown field, pushing the peak accelerating electric fields above 100 MV/m while protecting the cavity from dendritic thermomagnetic avalanches caused by local penetration of vortices.

\end{abstract}

\maketitle

\section{Introduction}
The maximum magnetic field $H$ which can be screened by a superconductor in the vortex-free Meissner state has attracted much attention, both from the fundamental and applied perspectives \cite{Bardeen,cavity,rast,qc,abrik,ste,exp1,exp2,exp3,exp4,aml,sh,shc,shd,posen,kubo}. This problem is particularly important for the Nb resonator cavities \cite{cavity} which have extremely high quality factors $Q(2K)\sim 10^{10}-10^{11}$ up to the breakdown fields $H_b\simeq 200$ mT at which the screening current density $J$ approaches the depairing limit \cite{Bardeen}, $J_d\simeq H_c/\lambda_0$, where $H_c\simeq 200$ mT is the thermodynamic critical field of Nb and $\lambda_0\simeq 40$ nm is the London penetration depth. The lack of radiation losses and vortex dissipation in the Nb cavities (clean Nb has the highest lower critical field $H_{c1}\simeq 180$ mT among type-II superconductors) enables one to probe the high-field nonlinear quasiparticle conductivity \cite{prl} which can be tuned by alloying the surface with impurities \cite{incr1,incr2,incr3}. 

The screening field limit of Nb could be exceeded by using s-wave superconductors with higher $H_c$ and the critical temperature $T_c$, but such materials are prone to the dissipative penetration of vortices at $H\simeq H_{c1}<H_{c1}^{Nb}$. To address this problem it was proposed to coat the Nb cavities with multilayers of thin superconductors (S) with high $H_c>H_c^{Nb}$ separated by dielectric (I) layers \cite{aml} (see Fig. 1). This approach is based on the lack of thermodynamically stable parallel vortices in decoupled S screens of thickness $d_s<\lambda$, which also manifests itself in a strongly enhanced $H_{c1}$ in thin films predicted by Abrikosov \cite{abrik,ste} and observed in Refs. \onlinecite{exp1,exp2,exp3,exp4}. 

The maximum field $H_m$ screened by $N$ superconducting layers of thickness $d=Nd_s\gg\lambda$ is limited by the superheating field $H_s$ of S-layers \cite{aml}, for example, $H_s\simeq 0.84H_c=454$ mT for Nb$_3$Sn at $T\ll T_c$. Here the Meissner screening  currents at $H=H_s$ become unstable with respect to infinitesimal perturbations of electromagnetic field and the order parameter, while the magnetic barrier for penetration of vortices vanishes \cite{sh,shc,shd}.  This paper addresses the limits to which the maximum screening field can be increased by S layers with given $d_s$, $\lambda$ and $H_s$ deposited onto a thick Nb substrate with given $\lambda_0$ and $H_{s0}$. It is shown that: 1. The maximum $H_m$ can be reached at an optimum multilayer thickness $d_m$ which depends on the materials parameters of S layers and the substrate; 2. The optimized S-I-S multilayer can screen the field exceeding both superheating fields, $H_s$ and $H_{s0}$; 3. S-I-S multilayers arrest thermomagnetic avalanches caused by local penetration of vortices at defects and do not let them develop into global flux jumps, which otherwise quench the cavity at fields well below $H_m$.     

\begin{figure}
\includegraphics[width=6.1cm]{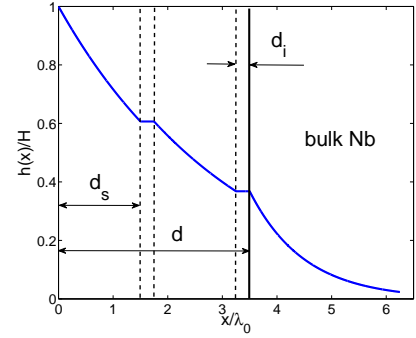}
\caption{\label{fig:one} Distribution of the magnetic field $h(x)$ in a multilayer of thickness $d$. The dashed lines show thin dielectric layers $(d_i\ll d_s)$ separating the superconducting layers. }
\end{figure}

\section{Superheating field, depairing current density and instability of Meissner state}

Distribution of a low-frequency $(\hbar\omega\ll k_BT_c)$ rf magnetic field $h(x)$ in a multilayer can be described by the London equations 
$\lambda^2h_1''=h_1$ at $0<x<d$ and $\lambda_0^2h_2''=h_2$ at $x>d$, where the prime denotes differentiation with respect to $x$.   Given that $h(x)$ in a multilayer calculated numerically from the Eilenberger equations is close to the result of the London theory \cite{shc}, I first disregard the nonlinear Meissner effect \cite{nme} which makes $\lambda$ dependent on $J$ at $J\sim J_d$.  Solutions of the London equations are (see also Refs. \onlinecite{posen,kubo}):
\begin{gather}
h_1(x)=H[(1-c)e^{-x/\lambda}+ce^{x/\lambda}], \quad 0<x<d,
\label{e1}
\\
h_2(x)=bHe^{(d-x)/\lambda_0}, \qquad x>d,
\label{e2} 
\\
c=\frac{k}{k+g^2},\qquad\quad b=\frac{(1+k)g}{k+g^2}.
\label{const}
\end{gather}
Here $h(x)$ and the rf electric field $E=-\mu_0\omega\lambda^2 h'$ 
are continuous at $x=d$, $k=(\lambda - \lambda_0)/(\lambda+\lambda_0)$, and $g=\exp(d/\lambda)$, both for a single S film and for a stack of S-I layers with $d_i\ll d_s$. 

For a S-I bilayer, the low-field surface resistance $\tilde{R}_s$ is determined by the total Joule rf power, $H^2\tilde{R}_s/2=(\sigma/2)\int_0^dE_1^2(x)dx+(\sigma_0/2)\int_d^\infty E_2^2(x)dx+q_i$, where $\sigma$ and $\sigma_0$ are the quasiparticle conductivities in S layers and the substrate, respectively, and $q_i$ accounts for dielectric losses. Using Eqs. (\ref{e1})-(\ref{const}), and $R_s=\mu_0^2\omega^2\sigma\lambda^3/2$ in the local dirty limit\cite{rast} yields \cite{err}
\begin{gather}
\tilde{R}_s=\frac{R_s}{(k+g^2)^2}\bigl[(g^2+k^2)(g^2-1)-4dkg^2/\lambda\bigr] \nonumber \\
+\frac{R_{s0}g^2(1+k)^2}{(k+g^2)^2}+\frac{4\mu_0^2d_i\omega^3\epsilon''\lambda^2\lambda_0^2g^2}{(\lambda+\lambda_0)^2(k+g^2)^2},
\label{rs}
\end{gather}
where $R_s$ and $R_{s0}$ include the residual resistances \cite{rast}, and $\epsilon=\epsilon'-i\epsilon''$ is the complex dielectric permeability of a low-loss I layer with $\epsilon''\ll \epsilon'$.

For $\lambda>\lambda_0$, the constant $c$ in Eqs. (\ref{e1})-(\ref{const}) is positive so   
the current density $J(0)=h_1'(0) = (1-2c)H/\lambda$ at the surface is lowered by the counterflow induced by the substrate \cite{kubo}. The Meissner state in S layers and the substrate is stable with respect to infinitesimal perturbations \cite{sh,shc,shd} if the current densities are smaller than the respective depairing limits, $J(0)=h_1'(0)\leq H_s/\lambda$ and $J(d)=h'_2(d)\leq H_{s0}/\lambda_0$. These conditions define the field region of the vortex-free state: 
\begin{equation}
H_s\geq\frac{H(g^2-k)}{g^2+k}, \qquad H_{s0}\geq\frac{Hg(1+k)}{g^2+k}.
\label{ineq}
\end{equation} 
Shown in Fig. 2 is the $H-d$ diagram where the Meissner region is stable below both curves defined by Eqs. (\ref{ineq}).  
If $H_s > H_{s0}\lambda_0/\lambda$, thin S layers with $d\ll \lambda$ are stable but do not screen the magnetic field which is thus limited by $H_{s0}$ of the substrate. As $d$ increases, 
$H(d)$ increases until $d$ reaches the crossing point $m$ at which the maximum field $H_m$ exceeds both $H_s$ and $H_{s0}$.  The latter results from the stabilizing effect of counterflow on $J(0)=(1-2c)H/\lambda$, where $c(d)$ given by Eq. (\ref{const}) decreases as $d$ increases.  As a result, $H(d)$ decreases with $d$ at $d>d_m$, approaching the superheating field of the S layer at $d\gg\lambda$. The optimal thickness $d_m$ is defined by the equalities in Eq. (\ref{ineq}) which give a quadratic equation for $g=\exp(d_m/\lambda)$. Hence,      
\begin{equation}
d_m=\lambda\ln \bigl( \mu +\sqrt{\mu^2+k} \bigr),
\label{dm}
\end{equation}
where $\mu=H_s\lambda/(\lambda+\lambda_0)H_{s0}$. Substituting $d_m$ back to one of Eqs. (\ref{ineq}) yields the maximum screening field
\begin{equation}
H_m=\left[ H_s^2+\left( 1-\frac{\lambda_0^2}{\lambda^2}\right)H_{s0}^2 \right]^{1/2}.
\label{hm}
\end{equation}
At $H=H_m$ and $d=d_m$, both current densities at $x=0$ and $x=d$ reach the depairing limits for the respective materials, but  $H_m$ in Eq. (\ref{hm}) exceeds both $H_s$ and $H_{s0}$ if $\lambda > \lambda_0$, because it is 
the current density $J\simeq J_d$ but not the magnetic field which makes the Meissner state unstable \cite{Bardeen,sh,shc,shd}. The Meissner region lying below both curves in Fig. 2 is similar to that was calculated numerically in the London model assuming that the energy barrier for the vortex parallel to the surface vanishes \cite{posen,kubo}. The London approach\cite{abrik,ste} used in Refs. \onlinecite{posen,kubo} has the well-known limitations, particularly at the fields $H\approx H_s$ for which the Bean-Livingston barrier is only a few vortex core diameters away from the surface. As a result, the London model cannot be used for the calculations of the superheating field, so the calculations of Refs. \onlinecite{posen,kubo} only give the right order of magnitude for $H_m$ but can hardly be used for quantitative evaluation of $H_m$.

Equations (\ref{ineq}) obtained from the condition of instability of Meissner currents have no ambiguity resulting from the evaluation of the surface barrier in the London model with the rigid core cutoff \cite{posen,kubo} which fails at $H\approx H_s$ because of strong deformation of the vortex core at the surface \cite{GL}. Calculations based on the Ginzburg-Landau (GL) theory have shown that the instability at $H=H_s$ is caused by lateral perturbations with the wavelength $\sim \xi^{3/4}\lambda^{1/4}$ decaying over the length $\sim\sqrt{\lambda\xi}$ perpendicular to the surface, where $\xi$ is the coherence length \cite{sh}. At $T\approx T_c$, the theory gives $H_s=f(\kappa)H_c$ where $f(\kappa)$ decreases as the GL parameter $\kappa=\lambda/\xi$ increases \cite{sh}, so that $H_s\approx 1.2 H_c$ at $\kappa\approx1$ and $H_s=(1+1/\sqrt{2\kappa})\sqrt{5}H_c/3$ at $\kappa\gg 1$.  At $T\ll T_c$, only the limiting value $f(\kappa)\to 0.84$ for $\kappa\gg 1$ has been calculated \cite{shc,shd}.  

\begin{figure}
\includegraphics[width=6.3cm]{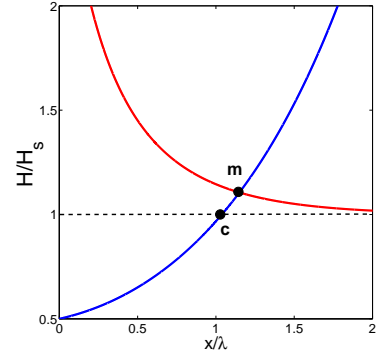}
\caption{\label{fig:two} Magnetic phase diagram of a multilayer. Meissner state is stable below both solid lines calculated from Eqs. (\ref{ineq}) for $k=1/2$ and $H_{s0}=0.5H_s$. 
Here $H(d)>H_s$ at $d>d_c=\ln(\mu+\sqrt{\mu^2-k})$, and the crossing point $m$ defines the optimum thickness $d_m$ at which $H(d)$ is maximum.  }
\end{figure}
  
Taking $\lambda=120$ nm and $H_s=0.84 H_c\simeq 454$ mT in Eq. (\ref{hm}) for a dirty Nb$_3$Sn (Ref. \onlinecite{nb3sn}) on Nb with $H_{s0}=240$ mT and $\lambda_0=40$ nm, yields $H_m=507$ mT at $d_m= 1.1\lambda= 132$ nm. Here the Meissner state persists up to the field $\simeq 12\%$ higher than $H_s$ of Nb$_3$Sn, consistent with the  London model of Ref. \onlinecite{posen}. For NbN films \cite{nbn} with $\xi = 9$ nm and $\lambda = 180$ nm, we obtain $d_m\simeq 79$ nm and $H_m\simeq $ 288 mT, while for  pnictides\cite{as,lambda} with $\lambda\simeq 200$ nm and $H_c\simeq 1$ T,  Eqs. (\ref{dm})-(\ref{hm}) give $H_s = 0.84H_s\simeq 840$ mT and  $H_m \simeq 872$ mT at  $d_m= 1.78\lambda = 356$ nm. The case of the two-band superconductor MgB$_2$ with $H_c \simeq 230$ mT is more complicated as the rf dissipation can occur at a smaller field $H_d\simeq H_c\xi_{\sigma}/\xi_{\pi}$ at which the screening current decouples two weakly coupled $\sigma$ and $\pi$ bands \cite{tb}. For the typical ratio of coherence lengths, $\xi_{\sigma}/\xi_{\pi}\simeq 0.2-0.3$, the band decoupling field $H_d\sim 50$ mT is consistent with the rf breakdown field of 42 mT at 4 K observed on MgB$_2$/Al$_2$O$_3$ bilayers on the Nb substrate \cite{exp2}.     

The enhancement of $H_m$ does not necessarily require I layers because Eqs. (\ref{dm})-(\ref{hm}) also describe a dirty layer at the surface 
where $\lambda$ is increased due to a shorter mean free path $\ell$.  For instance, a dirty Nb layer with $\ell\simeq 2$ nm has $\lambda\simeq\lambda_0(\xi_0/\ell)^{1/2}\simeq180$ nm and $\xi=(\ell\xi_0)^{1/2}\simeq 9$ nm.  Using $H_s\approx 0.84 H_c$ for $\kappa=\lambda_0/\ell=20$, yields $d_m=0.44\lambda = 79$ nm 
and $H_m=1.44 H_c = 288$ mT, the same as for the above case of NbN and some $20\%$ higher than $H_{s0}=240$ mT of pure Nb. In the limit of $(\lambda_0/\lambda)^2\to 0$, Eq. (\ref{hm}) 
gives $H_m=\sqrt{H_s^2+H_{s0}^2} = 1.465H_c = 293$ mT. 

To see how the nonlinear Meissner effect can change Eq. (\ref{hm}),  $H_m$ is also calculated from the GL theory at $\kappa\gg 1$ for which the $y$ component of the magnetic vector potential $A(x)$ satisfies the equation \cite{sh,shc,shd}
\begin{equation}
\lambda^2A''-A+\xi^2A^3=0.
\label{gl}
\end{equation} 
Consider a dirty layer with constant $\lambda$ and $\xi$ at $0<x<d$, a substrate with constant $\lambda_0$ and $\xi_0$, and the same $T_c$ in both regions. The first integrals of Eq. (\ref{gl}) are: $\lambda^2A'^2-A^2+\xi^2A^4/2=C$ at $0<x<d$ and $\lambda_0^2A'^2-A^2+\xi_0^2A^4/2=0$ at $x>d$. Continuity of $A(x)$ and $A'(x)$ at $x=d$, the boundary condition $A'(0)=-H_m$, and the GL pairbreaking conditions $A(0)=\phi_0/2\pi\sqrt{3}\xi$ and $A(d)=\phi_0/2\pi\sqrt{3}\xi_0$ under which the Meissner superflow becomes unstable \cite{sh,shc,shd} yield $(2\pi \lambda H_m/\phi_0)^2-5/18\xi^2=C_1$ at $x=0$, and $[2\pi h(d)\lambda/\phi_0]^2-1/3\xi_0^2+\xi^2/18\xi_0^4=C_1$ and $[2\pi h(d)\lambda_0/\phi_0]^2=5/18\xi_0^2$ at $x=d$. Here $C_1=(2\pi/\phi_0)^2C$, and $\phi_0$ is the magnetic flux quantum. Excluding $C_1$ and $h(d)$ and solving for $H_m$ reproduces Eq. (\ref{hm}) in which $H_s\to\tilde{H}_s$, and 
\begin{equation}
\tilde{H}_s^2=\left(1-\frac{\xi^2}{5\xi_0^2}+\frac{\xi^4}{5\xi_0^4} \right)H_s^2.
\label{hsm}
\end{equation}   
Here $H_s=\sqrt{5}H_c/3$, and both $T_c$ and $H_c=\phi_0/2^{3/2}\pi\lambda\xi=\phi_0/2^{3/2}\pi\lambda_0\xi_0$ are independent of the mean free path, according to the Anderson theorem \cite{shd}.  Two last terms in the parenthesis of Eq. (\ref{hsm}) resulting from the nonlinear Meissner effect give a negligible contribution to $H_m$ for a dirty layer with $\xi < \xi_0$.  
 
\section{Penetration of vortices at defects and thermomagnetic stability of multilayers.} 
 
The maximum screening field $H_m$ at which the Messier state becomes absolutely unstable with respect to infinitesimal perturbations can hardly be reached under realistic  operating conditions which require that the accelerating resonator cavities remain stable with respect to penetration of vortices, strong transient electromagnetic perturbations of charged beams, and local field enhancement at surface defects. In the multilayer approach \cite{aml} I layers are instrumental to assure the necessary stability margin with respect to local penetration of vortices at surface defects, which can otherwise trigger thermomagnetic flux jumps \cite{comm,aval} particularly at low temperatures $T\ll T_c$ and extremely high screening currents $J\sim J_d$ at which the cavities operate. Misinterpretation of this issue has lead to speculations that neither I layers nor the enhancement of $B_{c1}$ is necessary, so a few $\mu$m thick Nb$_3$Sn film coating of the Nb cavities could just be protected against penetration of vortices by the Bean-Livingstron surface barrier \cite{posen}. This assumption contradicts a vast body of experimental data on magnetization of high-$\kappa$ type-II superconductors for which inevitable materials or topographic defects at the surface cause premature local penetration of vortices at $H_{c1}<H < H_{s}$, or even $H<H_{c1}$ due to grain boundary weak links \cite{gb}. 

The maximum screening field $H_m$ at which the Meissner state is stable with respect to penetration of vortices can be evaluated from Eqs. (\ref{dm})-(\ref{hm}) with effective $H_s$ and $H_{s0}$ depending on the operating conditions. For instance, in a multilayer with $h(d) < H_{c10}\simeq 180$ mT at $H=H_m$, a vortex entering through a defect in the S layer cannot penetrate further into the bulk Nb. Let a surface defect cause local penetration of vortices as the current density, $J(0)=H'(0)=\beta H_{s}/\lambda$ reaches a fraction $\beta\lesssim 1$ of $J_d$  in S layer with $d_s<\lambda$ and $\xi\ll d_s$. If $J(0)$ is not too close to $J_d$, the  London theory shows \cite{ste,aml} that the energy barrier $U=\epsilon_0\ln(1/\beta)$ per unit length of a vortex in the S layer at $J=\beta J_d$ coincides with the bulk Bean-Livingston surface barrier \cite{abrik,ste} at $H=\beta H_{sh}\gg H_{c1}$ and $(\xi/d_s)^2\ll 1$, where $\epsilon_0=\phi_0^2/4\pi\mu_0\lambda^2$.  The criterion $J(0)<J_d/2$ assures a reasonable protection against penetration of vortices caused by low-angle grain boundaries in polycrystalline Nb$_3$Sn or pnictides \cite{gb}, or local field enhancement at typical topographic defects in the Nb cavities \cite{cavity}. 

At $H_s\to H_{s}/2$ and $H_{s0}\to 170$ mT,  Eqs. (\ref{dm})-(\ref{hm}) give $H_m = 278$ mT at $d_m=  0.8\lambda = 96$ nm for Nb$_3$Sn. If Nb can withstand the field $H_{s0}\to 200$ mT observed on the best cavities\cite{cavity,rast}, the maximum screening field could reach $H_m = 295$ mT at $d_m= 0.67\lambda = 80$ nm.  Increasing $\beta$ by the materials refinements of Nb$_3$Sn could push the peak fields over 300 mT. Pnictides with $H_c\simeq 0.9$ T, such as Ba$_{0.6}$K$_{0.4}$Fe$_2$As$_2$ (Ref. \onlinecite{as}) could provide $H_m = 426$ mT at $d_m = 1.21\lambda = 242$ nm, $\beta=1/2$, and $H_{s0}\rightarrow 200$ mT.  

\begin{figure}
\includegraphics[width=6.1 cm]{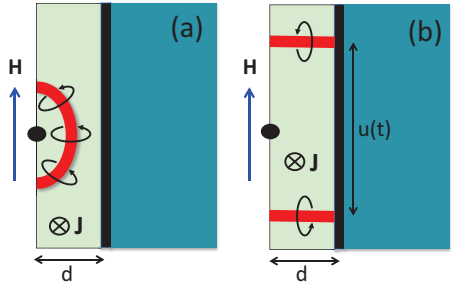}
\caption{\label{fig:three} Propagating vortex loop (a) turning into the vortex-antivortex pair (b) in S layer with a surface defect (black dot) which lowers the field threshold of vortex penetration.}
\end{figure}

To inhibit dissipative penetration of vortices, S film can be subdivided by I layers into $N$ thinner S layers with $d_s=d_m/N$.  At  $h(d)<H_{c10}$, even if a vortex penetrates at a defect in the first S layer, it could not propagate into the next S layer and further in the bulk Nb where it can cause a thermomagnetic avalanche \cite{aval}. As $H(t)$ reaches the critical value $\beta H_s$ at a week spot, a vortex line cannot penetrate parallel to the surface but first originates at a defect as a small semi-loop as depicted in Fig. 3.  The vortex semi-loop expands under the action of the perpendicular Meissner current until it hits the I layer where most part of the dissipative vortex core disappears in a loss-free flux channel connecting two short vortices of opposite polarity.  Because of the magnetic flux compression in the I layer and the substrate, the energy of a perpendicular vortex $\simeq d_s\tilde{\epsilon}_0\ln(L/\xi)$ diverges with the lateral film size $L$, while the energy of the vortex-antivortex pair $\simeq d_s\tilde{\epsilon}_0\ln(u/\xi)$ grows with the distance $u$, where $\tilde{\epsilon}_0\sim\epsilon_0$ (Ref. \onlinecite{vgk}). This vortex-antivortex pair expands during the positive rf cycle and contracts and annihilates as $H(t)$ changes sign. The upper limit of the pair size $u_m$ can be estimated neglecting the long-range vortex-antivortex attraction described by the last term in the dynamic equation,  $\eta\dot{u}/2=\phi_0J(t) - \tilde{\epsilon}_0/u$, where $\eta=\phi_0^2/2\pi\xi^2\rho_n$ is the viscous drag coefficient, and $\rho_n$ is the normal state resistivity.  Hence, $u_m\sim 2\phi_0 H/\lambda\eta\omega\sim\beta f\rho_n/\sqrt{2}\pi\kappa\mu_0\lambda\nu$ at $H=\beta H_s$ and $\omega=2\pi\nu$, giving $u_m\simeq 4\; \mu$m and the rf power \cite{gc}  $q\sim dJ^2\phi_0^2/\eta\sim \beta^2f^2d\phi_0B_c\rho_n/\kappa\mu_0^2\lambda^2 \sim 2\;\mu$W for Nb$_3$Sn at $\rho_n=0.2\;\mu\Omega$m, $\lambda=100$ nm, $\beta=1/2$, $f=0.84$, $\kappa=20$, $\nu=2$ GHz and $d_s/\lambda=0.2$.  Taking into account attraction of antiparallel vortices in S layer reduces $u_m$ and $q$.  Penetration of vortex semi-loops at a defect appears more realistic than the model of a long vortex parallel to the surface for which the ill-defined notion of $H_{c1}=0$ associated with the magnetic flux trapped in I layer \cite{posen} is irrelevant to the rf dissipation \cite{comm}.    

\section{Discussion}

Localization of the rf power in a thin S layer inhibits expansion of multiple vortex loops in the bulk and stops dendritic thermomagnetic avalanches \cite{aval} that are particularly pronounced at the extremely high screening current densities $J\sim J_d$ and low temperatures $T\ll T_c$ in the materials like Nb$_3$Sn, NbN or pnictides with low $\rho_n^{-1}$ and thermal conductivity $k$. The multilayer thus significantly reduces vortex dissipation as compared to the bulk Nb$_3$Sn, yet a thin Nb$_3$Sn layer with $d\sim 100$ nm may only slightly increase the thermal impedance of the cavity wall, $G=\alpha_K^{-1} + d_s/k_s+d_i/k_i+d_{Nb}/k_{Nb}$, where $\alpha_K$ is the Kapitza thermal resistance. For $d_{Nb}=3$ mm, $k_{Nb}\simeq 20$ W/mK, $\alpha_K=2$ kW/m$^2$K, the Nb$_3$Sn multilayer with $d_s = 100$ nm, $k_s\simeq 10^{-2}$ W/mK, and Al$_2$O$_3$ dielectric layers with $d_i=4$ nm and $k_i=0.3$ W/mK (Ref. \onlinecite{ins}) increases $G$ by only $\simeq 5\%$. A thicker Nb$_3$Sn film with $d\simeq 2-3 \mu$m doubles $G$ and reduces the field of thermal quench \cite{rast}, in addition to the smaller $H_{c1}=(\phi_0/4\pi\lambda^2)(\ln\kappa + 0.5) <130$ mT of bulk Nb$_3$Sn with $\lambda > 65$ nm (Ref. \onlinecite{nb3sn}) as compared to $H_{c1}^{Nb}\simeq 180$ mT. Theoretically, I layers provide the strongest pinning of propagating vortices and stop them more efficiently than holes in thin films which have been used to terminate dendritic flux avalanches \cite{avalo}. 

The optimum number of S layers for particular materials is determined by a balance between reduced vortex dissipation and suppression of superconductivity at the S-I interfaces. Here $H_s$ of ideal S layers with $d_s> (\xi\lambda)^{1/2}$ remains the same as in the bulk \cite{sh}, contrary to the assertion \cite{posen} that $H_s$ is reduced at small $d_s$. This claim was based on the artifacts of the London model discussed above and on taking into account only one right vortex image in Fig. 4 of Ref. \onlinecite{posen} instead of summing up an infinite chain of vortex-antivortex image dipoles which ensure that vortex currents do not cross the film surface. If this effect is properly taken into account \cite{ste,aml} $H_s$ in a thin film ($d < \lambda$) is the same as in a thick film ($d > \lambda$). Moreover, the GL simulations of Ref. \onlinecite{posen} for a Nb$_3$Sn film on Nb show that $H_m(d)$ reaches the maximum $H_m(d_m)\simeq 1.08H_{s}$ at $d_m\simeq 1.15\lambda$ but remains larger than $H_{s}$ in the whole region $\lambda < d < 2\lambda$ for which the numerical results were presented. 

High-field rf performance of the Nb cavities can be boosted by depositing not only materials with higher $H_s$ but also alloyed Nb-I-Nb multilayers which can increase $H_m$ and benefit from a significant raise of $Q(H)$ with $H$ in a wide field region \cite{prl,incr1,incr2,incr3}. A polycrystalline Nb multilayer may be tuned by alloying and heat treatment to reduce the residual resistance \cite{rast,prl,incr1}, and is also less prone to the current-blocking grain boundaries than the A-15 or pnictide compounds \cite{gb}.  Enhancement of the vortex penetration field by a dirty Nb/Al$_2$O$_3$ bilayer deposited onto the Nb cavity was observed in Ref. \onlinecite{nbml}. 

In conclusion, optimized multilayers can significantly increase the Meissner screening field while inhibiting dissipative penetration of vortices. Implementation of such multilayer coatings could potentially double the accelerating field gradients of superconducting resonators as compared to the existent high-performance Nb cavities.
   
This work was supported by DOE HEP under Grant No. DE-SC0010081.


\begin{references}
\bibitem{err}
 $\tilde{R}_s$ of Ref. \onlinecite{aml} reduces to Eq. (\ref{const}) at $\epsilon''=0$ and $\lambda=\lambda_0$, but  
is invalid at $\lambda\neq\lambda_0$ as the boundary condition of the continuity of 
$\lambda^2h'$ at $x=d$ was not taken into account \cite{kubo}.
\end{references}
\end{document}